 \documentclass[12pt,preprintnumbers,amsmath,amssymb,aip]{revtex4}
\usepackage{epsfig}
\usepackage{amssymb}
\usepackage{graphicx}

\input{epsf}
\begin{document}
\bibliographystyle{normal}

\draft
\title{Persistent global power fluctuations 
near a dynamic transition in electroconvection} 

\author{Tibor T\'oth-Katona$^{1,}${\footnote{On leave from: Research Institute for Solid State 
Physics and Optics, Hungarian Academy of Sciences, H-1525 Budapest, P.O.B.49, Hungary. E-mail: katona@physics.kent.edu}
John R. Cressman$^2$, Walter I. Goldburg$^2$ and James T. Gleeson$^1$
}
}

\address{$^1$Department of Physics, Kent State University, 
P.O.B. 5190, Kent, OH 44242\\
$^2$Department of Physics and Astronomy, University of Pittsburgh, Pittsburgh, 
PA 15260
}
\date{\today}

\begin{abstract}
This is a study of the global fluctuations in power dissipation and light 
transmission through a liquid crystal just above the onset of electroconvection.    
The source of the fluctuations  is found to be the creation and annihilation of defects.  
They are spatially uncorrelated and yet temporally correlated.    
The temporal correlation is seen to persist for extremely long times.
There seems to be an especially close relation between defect creation/annihilation in 
electroconvection and thermal plumes in Rayleigh-B\'enard convection.
\end{abstract}

\maketitle

Recently, the nature of fluctuations of global quantities such as power 
dissipation in systems held far from equilibrium has become of intense 
interest \cite{evans93,galla95}. Of particular importance is how fluctuating 
excitations on different spatial and temporal scales give rise to variations 
in globally-measured  quantities. 
Inspired by equilibrium systems, in which the relationship between local and 
global fluctuations is most easily studied near a 
phase transition, we examine  this relationship 
near a bifurcation point in a fluid dynamical system  
driven far from thermal  equilibrium. Here the classical fluctuation-dissipation  
theorem may not be invoked. 

Near such  a bifurcation only a small number of 
degrees of freedom are excited, making it possible 
to investigate both local and global quantities. It 
is well known that in weakly perturbed systems one 
can identify the correlation length of the system.  
Far from equilibrium, correlations in time are also 
highly relevant. An excellent system for studying 
temporal fluctuations in both global and local 
quantities is a liquid crystal in an electroconvective  state. 
This system permits straightforward, simultaneous, and temporally resolved measurement
of both localized spatial structures and global characteristics.
The principal bifurcation in electroconvection (EC) 
occurs above a critical driving voltage $U_c$ i.e., at 
the onset $\varepsilon \equiv (U/U_c)^2-1=0$ where convecting 
rolls appear.  A 
small increase in $U$ generates dislocations (defects) 
that translate across the plane of the liquid 
crystal (LC) and have a finite lifetime (see Fig. 1) leading to 
a state often called defect turbulence\cite{kai89}. 
These transient excitations cause readily measurable fluctuations 
in the global power $P(t)$.  The highly 
localized defects, which are tracked visually, have 
a lifetime which is in striking agreement with the 
correlation time of global quantities such as $P(t)$.      

In this Letter we show that the normalized variance of these fluctuations 
depends strongly on the system size, establishing 
that the local excitations (or quasi-particles) 
which give rise to the fluctuations in $P(t)$ are spatially 
uncorrelated. On the other hand, the temporal 
correlations in the system are very strong and show 
remarkable correspondence with the properties of 
thermal plumes in turbulent Rayleigh-B\'enard 
convection (RBC) \cite{ashken99a,qiu01a,qiu02}.  
The above observations are reconciled 
if one assumes that large-scale flow is responsible 
for the observed temporal coherence, while short-range 
interactions are responsible for their spatially 
incoherent distribution.  Our observations, made in three liquid crystals, may  apply to a large class of  
bifurcation phenomena, Rayleigh-B\'enard convection being a notable example.  

The experiments were performed on planarly oriented 
(nematic director parallel to the bounding glass plates) 
LC samples of methoxy benzylidene-butyl aniline (MBBA) 
Phase V (P5) \cite{merck} and Mischung 5 
(M5).\cite{jim02a}  All samples have been prepared with
appropriate electrical conductivities and the measurements were made under
temperature-controlled conditions.
The active area $A$ of the samples was varied between 
$A \approx 0.01 cm^2$ and $1cm^2$ while their thickness $d$ ranged between 
$(16.6 \pm 0.2) \mu m$ and $(52 \pm 1) \mu m$, providing aspect 
ratios $s=\sqrt{A}/d$ from $36$ to $602$. The details of 
the experimental setup for the electric power fluctuation measurements have 
been presented elsewhere \cite{jim01,jim02b,gold01}.  The set-ups permit  the recording of both $P(t)$ and the optical patterns, while at the same time monitoring the transmitted light intensity $I(t)$ integrated over the entire area of 
the sample. 

Fig. 2 shows the voltage dependence of the normalized variance in power 
fluctuations 
$\sigma_P/\langle P\rangle=\sqrt{\langle (P-\langle P \rangle )^2 \rangle}/\langle P\rangle$ 
measured in different cells filled with 
M5 and with various $A$ and $d$ providing 
$s$ in the range from 74 to 602. 
For all samples there is a sharp increase of 
$\sigma_P/\langle P \rangle$. The onset of this effect
is not $U_c$, but just above at
$\varepsilon \approx 0.2$ At this point
spontaneous creation and annihilation of defects begins and the stationary EC rolls 
break up into segments that move, seemingly at random with speed
proportional to $\varepsilon$ \cite{boden88}; this state has been called 
defect turbulence \cite{kai89}. 
Measurements in MBBA, not presented here, show that at any given value of 
$\varepsilon$ and $d$, $\sigma_P$ increases as $\sqrt{A}$, which
is proportional to the number of  defects. This strongly 
suggests that the  fluctuations in dissipation arises from spatially incoherent sources. 

Plotting the data in Fig. 2a in a different way suggests a relation between electroconvection and Rayleigh-B\'enard convection.  The inset in Fig. 2 is a plot of $\sigma_P/\langle P \rangle$, scaled by $s$, as a function  of $\varepsilon$.  Observe that the maximum in $(\sigma_P/\langle P \rangle)$ depends strongly on $s$.  When plotted in this way,
all curves collapse for $\varepsilon > 0.2$. This affirms that $s$ plays an important role here, as it does for velocity and temperature fluctuations in RBC (see e.g. \cite{qiu01a}).  Similar measurements  in MBBA  and P5 are found to exhibit the same behavior, suggesting that this is a     
 generic phenomenon in EC.

The non-monotonic  variation of  $\sigma_P/\langle P \rangle$ with $\varepsilon$ is 
understandable.  When $\varepsilon$ is just above zero, the dislocations, 
which are responsible for the fluctuations in $P$, are not yet excited. As $\varepsilon$ increases above
$\sim$ 0.2, the number of dislocations increases.  In this 
range the small-scale dislocations, seen in Fig. 1, possess a high degree of temporal coherence because 
the process of creation and annihilation is  stimulated  collectively by 
a circulation having  a size of the order of the lateral dimensions  
$L$ of the sample.   In the interval 0.2 $\alt \varepsilon \alt 1$,  
the localized dislocations   grow in number and their  motion  remains  spatially correlated, 
thereby increasing $\sigma_P$.  As for the denominator, $\langle P \rangle$,  it increases only  modestly 
\cite{gold01}.  Hence, the ratio $\sigma_P/\langle P \rangle$ increases in this  excitation range.  
As $\varepsilon$ is further increased, the number of dislocations continues to grow, but now the 
large-scale spatio-temporal  coherence is lost; the fluid motion  has become strongly chaotic on all scales.  
As a consequence, the defects contribute  randomly to the fluctuations in  $P$, causing   
$\sigma_P/\langle P \rangle$ to fall
According to this scenario, the strong dimensionless fluctuations in $P$ (and very likely, other global quantities such as $I(t)$)will be confined to a narrow interval in $\varepsilon$ near zero.

Fig. 3(a) shows time traces of both $P(t)$ 
 and the integrated transmitted intensity $I(t)$ 
 at $\varepsilon=3.0$ and at $\varepsilon=0.78$. 
Both of these global measures are quasi-periodic with a dominant 
frequency $f^*$ that increases with $\varepsilon$. 
The relative phase between the signals is arbitrary.
Note  that low frequency oscillations in $I(t)$ were seen in Ref.
\cite{kai78}; no other reported examples are known to us.

This characteristic frequency $f^*$, is also seen in the normalized autocorrelation function 
$g_a(t)=\langle P(t^{\prime})P(t^{\prime}+t)\rangle / \langle P \rangle ^2  - 1$ displayed in Fig. 3(b). 
The measurements in 
Fig. 3(b) were made in an M5 sample 
($A=50mm^2$, $d=52\mu m$) for different values of $\varepsilon$. 
At $\varepsilon=0.124$, where EC rolls are well developed but there 
is no defect creation/annihilation, $g_a(t)$ reaches $0$ within a minute 
without any oscillation. When the so called 'varicose' 
\cite{nasuno92} pattern develops in M5 at 
$\varepsilon \approx 0.2$, the motion of the rolls starts with infrequent
generation/annihilation of defects, and slow oscillations  at frequency $f^*$.  With further increase of $\varepsilon$, $f^*$ increases up to $\varepsilon \approx 5$, 
where so-called hard turbulence 
takes place with rapid creation/annihilation of defects. 
Above $\varepsilon \approx 5$ the oscillations vanish as
their amplitude disappears into the noise.

In the range of $\varepsilon$ for which oscillations
in $g_a(t)$ are seen, they appear to be truly persistent.
That is, even for measurement times of many hours, the
oscillations persist for the length of the run.
A run of this duration spans tens of thousands of oscillation periods.  
These measurements were at a value of $\varepsilon$ where the sample 
contained hundreds of defects, if not thousands.  It  is surprising 
that such a large number of  spatially  uncorrelated defects  gives 
rise to oscillations in  $g_a(t)$ over such long times.

The $\varepsilon$ dependence of $f^*$ obtained from $g_a(t)$ in M5 samples 
is presented in Fig. 4 for four different aspect ratios (solid symbols).
The open circles in this figure are 
inverse lifetimes $\tau^{-1}$, as observed from visually tracking the evolution of defects, 
a few of which are seen in Fig. 1.  
Each such data point is from $\tau$ averaged over up to ten defects.  
Stunningly, the inverse lifetimes are, within typically 5\%, 
the same as $f^*$.   We submit that this close agreement leaves little doubt that
defects are the localized excitations responsible for
fluctuations in the global, dissipated power.  
The inset of Fig. 4  better shows that the oscillations in $g_a(t)$ start 
at $\varepsilon \approx 0.2$-well above the onset of EC.

Figure 1  displays  a series of photographs made at the indicated times  
$t$. 
The liquid crystal is M5 having a thickness, $d=52 \mu$m.  In this relatively  thick  sample,  the 'varicose' pattern persists down  to $\varepsilon = 0.2$
At this low level of  excitation,  the number of generated defects is relatively small and the motion of 
the rolls is slow. At $t=0$ we begin watching a single 
defect of interest (marked with a black circle) as it starts to annihilate 
[Fig. 1(a)]; $1s$ later the annihilation process is finished [Fig. 1(b)] and 
a defect is created at the same location about $26s$ later [Fig. 1(d)-(e)]. 
Note, there is a slight difference in the position of the defects in 
Fig. 1(a) and (e) due to the slow climb motion \cite{nasuno89}. 

Optically tracking dislocations 
is problematic as the number of defects increases and their lifetime
decreases.  The frame grabber has limited time resolution, and 
when the dislocations are too dense, it becomes difficult
to track individual dislocations as the 
climb/glide motion of the defects speeds up\cite{nasuno89}. 
Measuring power fluctuations has no such limitation. Thus,
having identified $1/f^*$ as the defect lifetimes, we are able
to determine this quantity beyond the range where it is possible
to do so optically.
In Fig. 4, $f^*$ is measured over the 
whole range of $\varepsilon$ 
where so-called weak turbulence (where the spatial coherence is 
destroyed)\cite{nasuno92})   
occurs up
to the hard turbulence transition (DSM1), where oscillations in $g_a$ 
diminish. 

In RBC \cite{qiu02} when 
the Rayleigh number $Ra$ is increased above a threshold,
the  rate of generation of thermal plumes increases, and, they
begin interacting via their self-generated flow field. 
This occurs in the same $Ra$ range in which the  
transition from soft to hard turbulence occurs. 
The plume motion is  temporally correlated,
producing coherent oscillations in both temperature and
velocity (as observed by monitoring a single point in space).
Villermaux's model\cite{viller95} predicts the frequency of these oscillations
should scale with $d^2$ and increase with $\sqrt{Ra}$.
Within this model, the oscillations arise from a  recirculation 
flow having length scale comparable to the entire system.

On the other hand, in EC the continuous generation/annihilation of dislocations 
(defect turbulence) results from an advection of the roll pattern by the 
mean flow, which amplifies 
small undulations in the director field \cite{kaiser93}. Because the boundary 
conditions counteract the bending of rolls, the stress is released by 
straightening the rolls and topological defects are left behind. 
A number of experimental studies have been devoted to the motion of defects 
and the process of their creation and annihilation \cite{nasuno92,nasuno89,rasen90}.
Most of those  measurements were  made at relatively low $\varepsilon$, where few defects are created.
In the context of the present work, the most important result of those 
studies is that two kinds of forces determine the motion of the defects: 
a large scale pattern-selection force and a short-range interaction force 
(for details see \cite{nasuno89}). 

The similarities between RBC thermal plumes and EC dislocations are striking. 
Both plumes and dislocations are generated in the presence of a large 
scale flow. They interact with the neighbouring plumes (dislocations) and 
they organize themselves in space and time producing coherent oscillations. 
The most convincing evidence connecting EC and RBC comes from the  agreement between
our experimental  results  and the prediction of Ref. \cite{viller95}.  There Villermaux 
shows, for RBC,  that 
the quantity $f^*$ should obey the relation, $f^*d^2 \propto \sqrt{Ra}$. 
Figure 5 shows  $f^*$ obtained from electric power fluctuation 
measurements ({\em c.f.} Fig. 4) scaled with $d^2$.  The abcissa is
$\varepsilon_d = \varepsilon - \rm{const}$, where $\varepsilon_d=0$ is defined 
as the threshold of the defect creation/annihilation (${\rm const} \approx 0.2$).\cite{Ra} 
All data for different $d$ (and $s$) 
collapse into the same curve. 

In conclusion, we have identified the source of the dramatic
increase of the normalized variance in dissipated power 
in electroconvection as coming from the spontaneous generation and annihilation of dislocations
in the roll pattern. Even though not spatially coherent, these localized excitations, of limited lifetime,
lead to persistent, long-time temporal coherence in  global quantities.
Similar coherence, seen in  Rayleigh-B\'enard convection, is associated with the
generation of thermal plumes.  From Fig. 5, and the calculations of Villermaux, we conclude that dislocations in 
EC play the role same role as do thermal plumes in RBC.

\begin{acknowledgements} 
This work us supported by the 
by the National Science Foundation under Grants DMR-9988614  and
DMR-0201805.
\end{acknowledgements}

\newpage
\begin{figure}[h]
\begin{center}
\parbox{16.5cm}{
\epsfxsize=16cm
\epsfbox{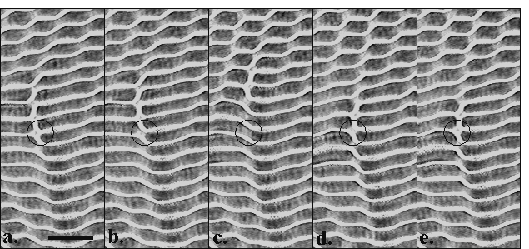}}
\end{center}
\caption{Snapshots showing the defect creation/annihilation process in time: 
$t=0$, $1s$, $15s$, $26s$, and $27s$ for subfigures (a)-(e), respectively. 
The circle mark points the same location in the cell ($s=136$, 
filled with M5, $\varepsilon=0.2$). The scale bar shows $100 \mu m$.
}
\label{fig1} 
\end{figure}

\newpage
\begin{figure}[h]
\begin{center}
\parbox{16cm}{
\epsfxsize=15cm
\epsfbox{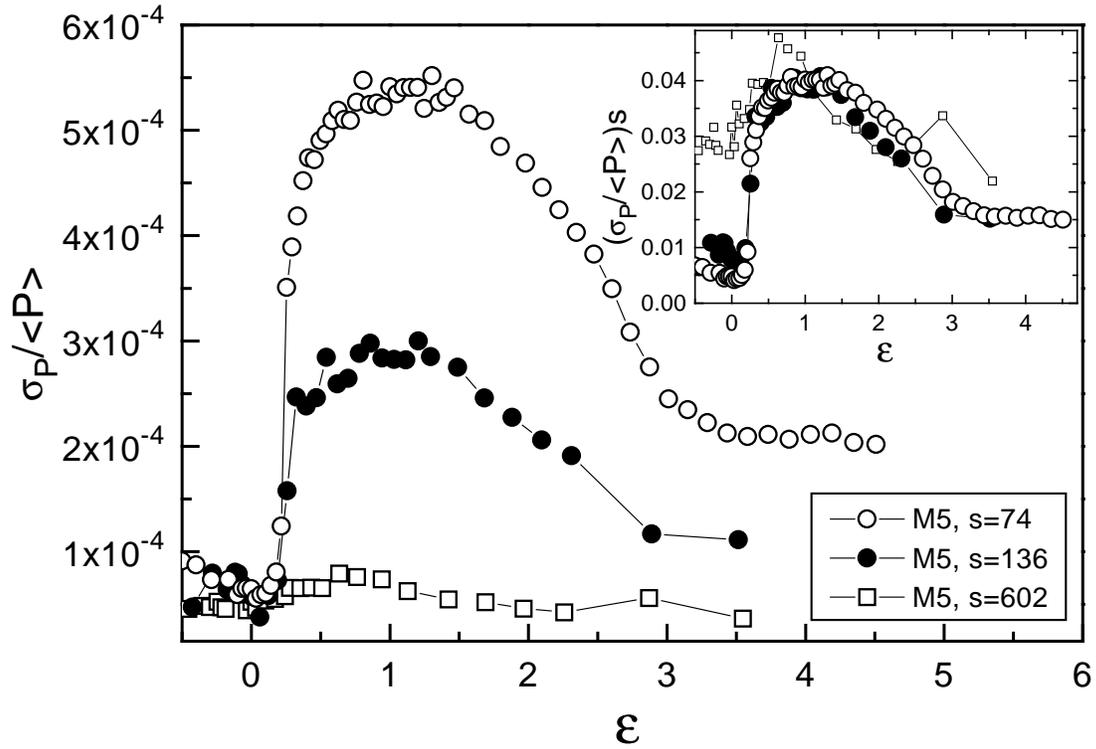}}
\end{center}
\caption{Dimensionless voltage dependence of the width $\sigma_P$ of the 
power fluctuations normalised with the mean value of power 
$\langle P\rangle$ measured in cells with different aspect ratio 
$s$ and filled with M5.
Inset: $\sigma_P/\langle P\rangle$ normalised with $s$ vs. 
$\varepsilon$.
}
\label{fig2} 
\end{figure}

\newpage

\begin{figure}[h]
\begin{center}
\parbox{15cm}{
\epsfxsize=14cm
\epsfbox{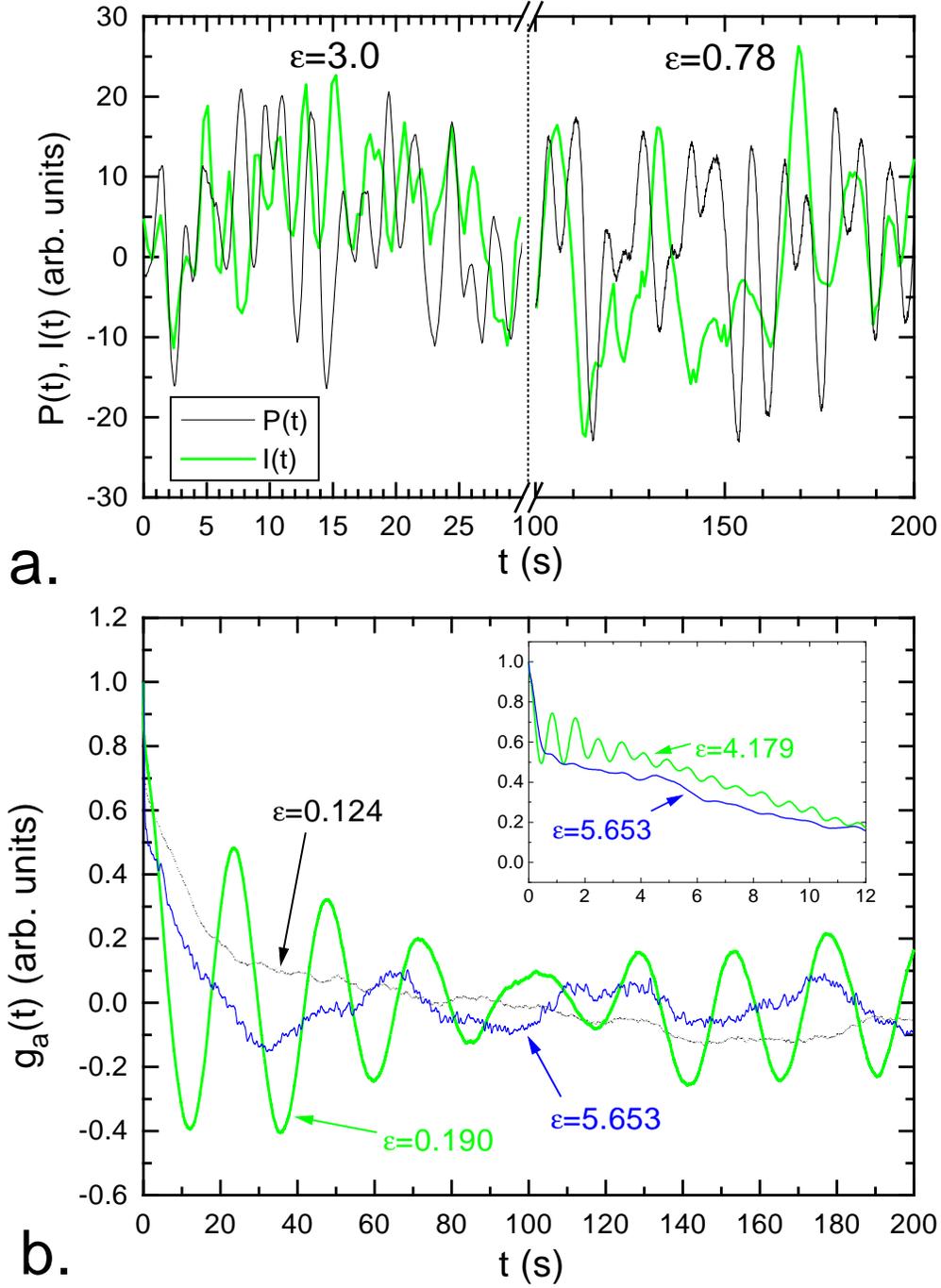}}
\end{center}
\caption{(a) Time series traces of both electric power $P(t)$ and 
transmitted light intensity $I(t)$ in an MBBA sample for $\varepsilon=3.0$ 
and $\varepsilon=0.78$ ($A=10mm^2$, $d=50\mu m$). The power and transmitted 
light intensity were measured at different times. 
(b) Autocorrelation function $g_a(t)$ measured in an M5 sample 
($A=50mm^2$, $d=52\mu m$) for different values of 
$\varepsilon$. Inset: the same as the main graph for relatively high 
values of $\varepsilon$ where the oscillations in $g_a(t)$ diminish.
}
\label{fig3} 
\end{figure}

\newpage

\begin{figure}[h]
\begin{center}
\parbox{16.5cm}{
\epsfxsize=16cm
\epsfbox{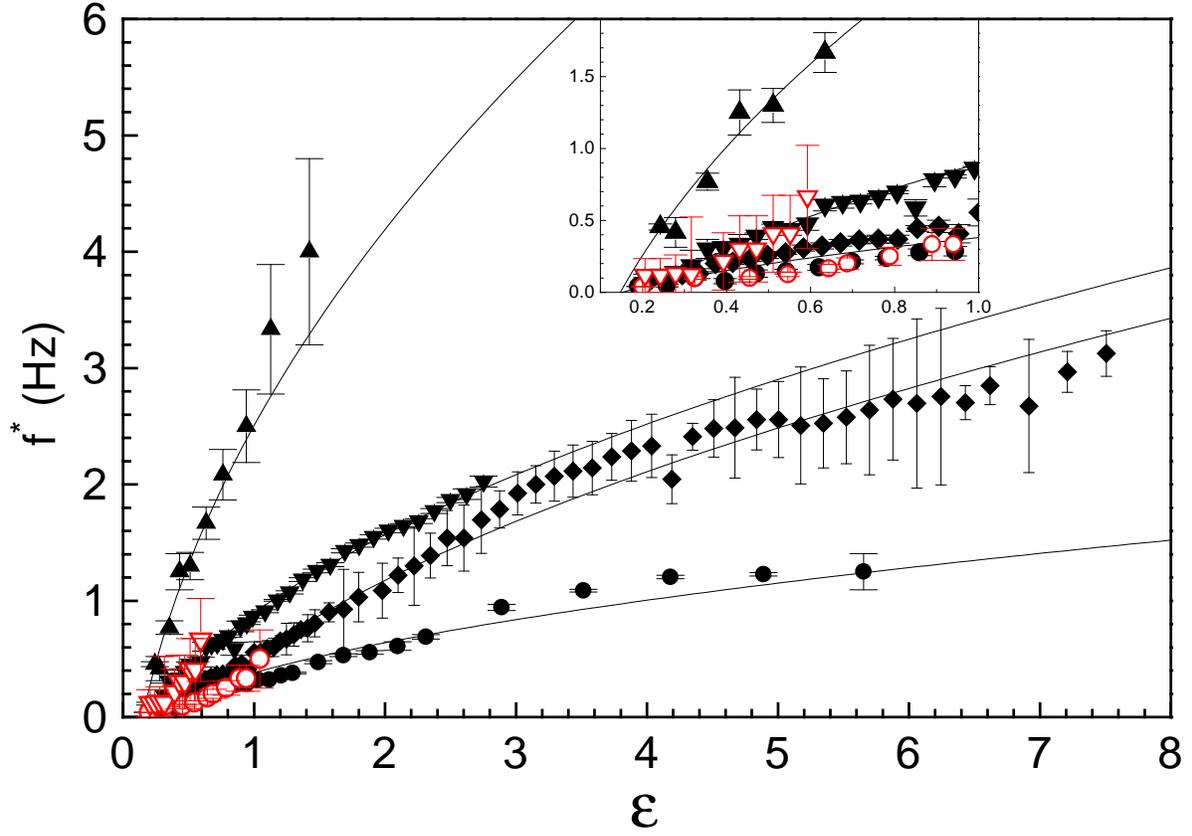}}
\end{center}
\caption{Frequency $f^*$ of oscillations extracted from $g_a(t)$ 
(filled symbols) and determined from the optical observation of the 
defect creation/annihilation rate (open symbols) as a function of driving 
voltage $\varepsilon$. Measurements have been performed on M5 samples with 
different aspect ratios: 68 (down triangles), 74 (diamonds), 
136 (circles) and 602 (up triangles). 
Solid lines are square-root fits to the data (see later 
discussion). Inset: the blow-up of the low $\varepsilon$ region. 
}
\label{fig4} 
\end{figure}

\newpage
\begin{figure}[h]
\begin{center}
\parbox{16.5cm}{
\epsfxsize=16cm
\epsfbox{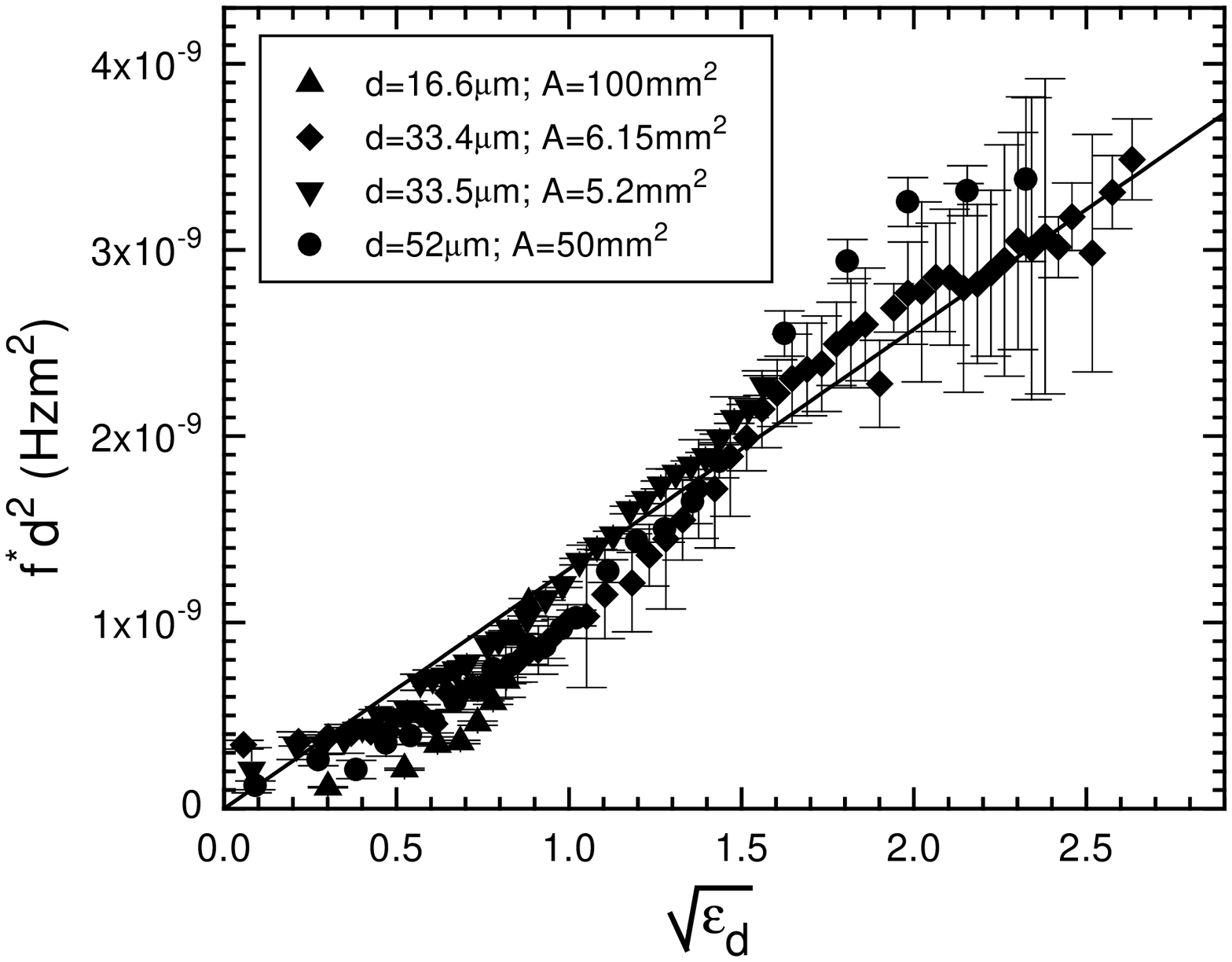}}
\end{center}
\caption{Frequency $f^*$ of oscillations extracted from $g_a(t)$ 
normalised with the sample thickness (see discussion) 
as a function of dimensionless square-root voltage $\sqrt{\varepsilon_d}$ 
in M5 samples. The solid line represents a linear fit to the data. 
}
\label{fig5} 
\end{figure}


\begin{references}
\bibitem{Ra}
$\varepsilon$ is properly compared directly with the reduced Rayleigh number, 
$r \equiv Ra/Ra_t ~-~ 1$, in which $Ra_t$ is the threshold for RBC.  Plumes are seen 
$r \gg 1$, where the only difference between $r$ and $Ra$ is a multiplicative constant.
\end{references}
\end{document}